\documentclass[16pt,aps,nofootinbib]{revtex4}
\usepackage{amsmath}
\usepackage{graphicx}
\usepackage{slashed}

\renewcommand{\vec}[1]{\boldsymbol #1}

\begin{document}

\title{Naive versus mirror assignment at finite density: correlator mixing and chiral symmetry restoration.}
\author{Boris Krippa$^{1}$}
\affiliation{$^1$Department of Physics, Imperial College London, London, UK}
\date{\today}
\begin{abstract}
Chiral symmetry imposes some constrains on hadron correlation functions in dense medium. These constraints imply a certain general structure of the two- and three-point correlation functions of chiral partners and lead to the  effect of mixing arising from the interaction of the long-ranged nuclear pions with corresponding interpolating currents. It reflects the phenomena of partial restoration of chiral symmetry. We consider several possible pairs of chiral partners for both mesons and nucleons and analyse their mixing patterns.  We explored both naive and mirror assignments for the chiral partner of nucleon.

\end{abstract}
\maketitle

Understanding of a phase structure of dense hadron matter is one of the key problems in the strong interaction physics described by quantum chromodynamics (QCD).  At low energy QCD can be reduced to an effective theory containing the Goldstone bosons as the effective degrees of freedom. The extension of this effective theory to the  finite density requires taking into consideration interactions of the bosons with the medium. The physical motivation for such an extension 
is provided by a wide range of phenomenon occurring in nuclear matter, heavy-ion collisions and the Equation of State  (EoS) of the neutron stars \cite{Lat}. 

The most straightforward way of calculating observables in QCD from the first principles is to use the lattice methods extended to the finite density case. 
However, a well known problem in such an extension has been the inability to perform lattice gauge theory ab initio simulations by standard Monte Carlo methods since the fermion determinant becomes complex in the presence of finite baryon chemical potential, the notorious “sign problem.” In spite of the significant progress which has been made in recent years in overcoming some of these problems most of the numerical studies in this area focus   on small chemical potential, semianalytical techniques, and investigation of model systems. However, away from the region along the temperature axis, the QCD phase diagram remains, for the most part, inaccessible for lattice methods.

 In particular, hadron dynamics remains to be understood at the densities where the system is believed to undergo phase transition related to the chiral symmetry restoration(CSR)phenomena. Therefore, developing effective models is nessesary and inevitable. Many such models have been proposed to describe different aspects of hadronic matter at finite density  including quark-meson  \cite{Sch} and dilaton models \cite{Mish}, chiral models with hidden local symmetry \cite{Moto} just to name a few.

All these models predict the onset of CSR at some critical chemical potential but the details of the phase transition like the corresponding critical temperature/density or the order of the transition differ  from model to model. Therefore, it is important to distinguish the model dependent artefacts from the genuine dynamical features. Therefore, the studies where the CSR related phenomena are analysed in a model independent way seem to be useful. Certainly such studies are admittedly limited but important for establishing constraints and conditions each realistic effective model of CSR should satisfy.

In this paper we analyse such model independent phenomena related to CSR, namely mixing of the  correlation functions, describing chiral partners. Such mixing arises in the hadronic matter, regardless of the concrete model assumed for the correlator and is due to the Goldstone bosons interacting with interpolating fields entering the correlator.

The paper is organised as follows. We will first describe some formal aspects related to chiral mixing of the hadron correlation functions in the medium and then discuss the properties of the nucleon two-point correlators both in the naive and mirror assignments as well as the in-medium behaviour of the quark condensate. The next section is devoted to an extension of the analysis to the case of the 3-point correlation functions and then we conclude with a summary and outlook.

The main object we focus on is the correlation function of certain hadron interpolating currents in nuclear medium. We first illustrate the technique using the case of the two-point correlation function and then consider three-point correlators.  A general correlation function can be written as 
\begin{equation}
\Pi(p) =-i \int d^4x \, exp(i p x)<\psi_A|T{J(x) J(0)}|\psi_A>
\end{equation}

Here $\psi_A$ is a full nuclear wave function and $J(x)$ is the interpolating field with the quantum numbers of interest. The nuclear wave function can schematically be written as
\begin{equation}
\psi_A = |\psi> + \sum_a |\psi\pi_a> + \sum_{a,b}|\psi\pi_a\pi_b> + ......... ,
\end{equation}
where the first term represents the system of bare noninteracting nucleons and the second and third terms corresponds to the system where one and two virtual pions are added to the system.
Using this expansion the correlator can be represented as follows
\begin{equation}
\Pi(p) = \bar{\Pi}(p) + \Pi_{\pi}(p),
\end{equation}
Here $\bar{\Pi}(p)$ and $\Pi_{\pi}(p)$ are the correlators of the infinite system of bare nucleons and with pion corrections correspondingly. After some rather straightforward algebra (details can be found in Ref. \cite{Kri1}) the correlator $\Pi_{\pi}(p)$ can be written as a sum of three terms

\begin{equation}
\Pi_{\pi}(p) = \Pi_1(p) + \Pi_2(p) + \Pi_3(p),
\end{equation}
where
\begin{equation}
\Pi_{1}(p) =  \sum_{a,b} \int \frac{d^3k d^3k'}{2\omega_{k}2\omega_{k'}} <\psi|a_{a}^{+}(k) a_{b}(k')|\psi> i \int d^4x \, exp (i p x) <\psi\pi_a(k)|T\{J(x) J(0)\}|\psi\pi_b(k')> ,
\end{equation}
\begin{equation}
\Pi_{2}(p) =   \frac{1}{2}\sum_{a,b} \int \frac{d^3k d^3k'}{2\omega_{k}2\omega_{k'}} <\psi|a_{a}^{+}(k) a_{b}^{+}(k')|\psi> i \int d^4x \, exp (i p x) <\psi\pi_a(k)\pi_b(k')|T\{J(x) J(0)\}|\psi> ,
\end{equation}
and
\begin{equation}
\Pi_{3}(p) =  \frac{1}{2}\sum_{a,b} \int \frac{d^3k d^3k'}{2\omega_{k}2\omega_{k'}} <\psi|a_{a}(k) a_{b}(k')|\psi> i \int d^4x \, exp (i p x) <\psi|T\{J(x) J(0)\}|\psi\pi_b(k')\pi_b(k')> ,
\end{equation}
where $\omega_k = \sqrt{k^2 + m^2_\pi}$.
The term $\Pi_1$ corresponds to the process where the interpolating current interacts with the in-medium pion which is first emitted and then absorbed by the matter. Another terms describe the other time orderings where the pions are either emitted or absorbed. The next step is to use the soft pion theorem which gives 

\begin{equation}
 <\psi\pi_a(k)|T\{J(x) J(0)\}|\psi\pi_b(k')> = -\frac{1}{f^2_{\pi}} <\psi|\left[Q_{5}^{a}\left[Q^{b}_{5},T\{J(x) J(0)\}\right]\right]|\psi> 
\end{equation}
Here $f_{\pi}$ and $Q_{5}^{a}$ are pion decay constant and axial generator correspondingly. The other terms can be transformed in the same way. Now the pionic part of the correlator takes the form
\begin{eqnarray}
\Pi_{\pi}(p) = \frac{1}{2} \sum_{a,b} \int \frac{d^3k d^3k'}{2\omega_{k}2\omega_{k'}}\int d^4x \, exp (i p x) \left( <\psi|2 a_{a}^{+}(k) a_{b}(k') + a_{a}^{+}(k) a_{b}^{+}(k') + a_{a}(k) a_{b}(k'))|\psi>\right) \nonumber \\
\left(-\frac{i}{f^2_{\pi}} <\psi|\left[Q_{5}^{a}\left[Q^{b}_{5},T\{J(x) J(0)\}\right]\right]|\psi>\right) ,
\end{eqnarray}
where the momenta $k$ and $k'$ in the creation and annihilation operators are assumed to be vanishing. The integral over momenta is related to the matrix element
$<\psi|\frac{1}{2}\pi^2(0)|\psi>$. To the first order in density this matrix element can be represented as
\begin{equation}
(2\pi)^3 m^2_{\pi}<\psi|\frac{1}{2}\pi^2(0)|\psi>\simeq (2\pi)^3 m^2_{\pi}<N|\frac{1}{2}\pi^2(0)|N> \rho\simeq \rho\bar{\sigma}_{\pi N},
\end{equation}
where $\rho$ is the nuclear density and $\bar{\sigma}_{\pi N}$ is the leading nonanalytic part in the chiral expansion of the pion-nucleon sigma term $\sigma_{N}$ resulting from the long distance contribution of the pion cloud so that it is natural that this term enters  the chiral expansion of the in-medium correlator. Following the chiral perturbation theory results \cite{Gas} the value of the nonanalytic term is taken to be -20 MeV.


 The actual form of the expansion depends on the commutation relation between the axial generator $Q^a_5$ and concrete interpolating field. Below we will consider  several physically relevant cases involving different types of the interpolating fields.
\begin{center}
1. CHIRAL MIXING OF NUCLEONS IN NAIVE AND MIRROR ASSIGNMENTS.
\end{center}
It is well known that current quark mass can explain about 2\% of the nucleon mass. The accepted mechanism of the nucleon mass generation is related to the spontaneously broken chiral symmetry with quark condensates as order parameters. At finite temperature/density chiral symmetry is believed to be partially restored so that according to the standard interpretation nucleon becomes light and degenerate with its chiral partner, belonging to the same representation of the chiral group.

There are two ways to introduce the chiral partner of the nucleon. One method utilizes the so called naive assignment \cite{Lee} when the mass of the nucleon is generated only via spontaneous chiral symmetry breaking as an explicit mass term would break chiral invariance. In the naive assignment the chiral transformations for nucleon $N$ and its chiral partner $N^{*}$ are defined as 
\begin{equation}
N_R \rightarrow U_R N_R, \qquad N_L \rightarrow U_L N_L , \qquad
N^*_R \rightarrow U_R N^*_R, \qquad N^*_L \rightarrow U_L N^*_L 
\end{equation}
Here $U_L$ and $U_R$ are matrices, indicating left and right chiral transformations.  In this assignment chiral symmetry requires that the nucleon and its chiral partner must be almost massless and degenerate in the Wigner phase. However, this degeneracy is rather trivial and is not related to the chiral symmetry between $N$ and $N^*$. It means that chiral symmetry puts no constraints on relation between nucleon and its chiral partner and implies that the off-diagonal coupling $N^*N\pi$ vanishes in the chiral limit.

In the second way which is called mirror assignment \cite{Kun} the transformation rules are 

\begin{equation}
N_R \rightarrow U_R N_R, \qquad N_L \rightarrow U_L N_L , \qquad
N^*_R \rightarrow U_L N^*_R, \qquad N^*_L \rightarrow U_R N^*_L 
\end{equation}

It allows for a chirally invariant explicit mass term in the effective Lagrangian which sirvives when chiral symmetry is restored. Similar to the naive assignment in the restored phase  two nucleon masses may become degenerate but unlike the naive assignment the mass inself stays finite. This scenario was realised in the parity doublet model \cite{Kun} which was then used to describe resonances \cite{Hat}, the properties of the baryon octet \cite{Nem} and extended to the finite density case \cite{Sasa}. 

We will focus on some physical consequences following from these two assignments in the case of nuclear matter. Let's first consider naive assignment. This case has been studied in details in Ref. \cite{Kri2} so here we just summarise the main line of reasoning. The effective mass of nucleon in dense matter gets modified due to scalar self-energy. Suppose it depends on $\bar{q}q$ condensate in some way. Then one can expect a piece in the self-energy which is linear in density and proportional to pion-nucleon sigma term $\sigma_{\pi N}$. At moderate densities the nucleon self-energy can be represented by the product of the scalar part of the nucleon-nucleon amplitude and nuclear density. Using chiral expansion of $\sigma_{\pi N}$ one can see that the nucleon-nucleon scattering amplitude contains a term linear in $m_\pi$, which is not allowed by chiral symmetry \cite{Wei}. Therefore, this term breaks chiral symmetry and should somehow be cancelled. As it was shown in the framework of the QCD sum rules (QCD SR) approach \cite{Coh} the unwanted pieces in Operator Product Expansion (OPE) can be cancelled by the certain piece of the phenomenological side of the sum rules explicitly containing contribution of the pion cloud. Similar cancellation happens in matter as well.

We start from the standard expression for the correlator of the nucleon interpolating fields

\begin{equation}
\Pi (p)=-i \int d^4x \, exp (i p x) <\psi|T\{\eta(x) \bar{\eta}(0)\}|\psi>,
\end{equation}
where the Ioffe's choice \cite{Ioff1} of the nucleon interpolating field is assumed.

In the naive assignment the commutation relation of the interpolating field with the axial charge takes the form
\begin{equation}
[Q^{a}_5, \eta (x)] = - \gamma_5 \frac{\tau^a}{2}\eta(x)
\end{equation}

Using this relation one can get the chiral expansion of the correlator

\begin{equation}
\Pi (p) = \bar{\Pi}(p) - \frac{\xi}{2} ({\bar{\Pi}(p)} + \gamma_5 \bar{\Pi}(p)\gamma_5),
\end{equation}
where $\xi = \frac{\rho \sigma_{\pi N}}{f^2_\pi m^2_\pi}$.

Similar expression can be obtained for the nucleon chiral partner

\begin{equation}
\Pi^* (p) = \bar{\Pi}^*(p) - \frac{\xi}{2} ({\bar{\Pi}^*(p)} + \gamma_5 \bar{\Pi}^*(p)\gamma_5),
\end{equation}

One can also obtain the expression for the difference $\Pi_N (p) - \Pi_{N^*} (p) $ which can be viewed as an order parameter. This order parameter could be useful when analysing restoration of chiral symmetry in the framework of QCD SR because of the multiple cancellations of rather poorly known in-medium quark condensate. 

\begin{equation}
\Pi_N (p)  - \Pi_{N^*} (p) = (\bar{\Pi}_N (p) - \bar{\Pi}_{N^*} (p))(1 - \frac{\xi}{2} ) - \frac{\xi}{2} (  \gamma_5(\bar{\Pi}_{N}(p) - \bar{\Pi}_{N^*} (p))\gamma_5),
\end{equation}

One notes that this expression does not show any tendency towards chiral symmetry restoration where the difference between the correlators of the chiral partners should decrease. On the contrary, both terms act against it so that the difference between the correlators grows with the increase of nuclear density. It contradicts the common belief that the correlators of the chiral partners should become degenerate with increasing density/temperature. As a possible consequence it may indicate the fact that the naive assignment is not a correct one. Certainly, higher order corrections need to be taken into account before any definite conclusion can be drawn.

The correlator can be decomposed into three terms with different Dirac structures \cite{Coh1}

\begin{equation}
\Pi (p) = \Pi^{s}(p) + \Pi^{p}(p)\slashed p + \Pi^{u}(p) \slashed u,
\end{equation}
where $u^{\mu}$ is a unit-4-vector defining the rest frame of the system. The terms proportional to $\rho m_{\pi}$ contribute to the piece $\Pi^{s}(p)$. Splitting the phenomenological part of the correlator into pole and continuum parts one can obtain

\begin{equation}
\Pi (p) \simeq \Pi_{pole}(p) - \frac{\xi}{2} \gamma_5 \Pi_{pole}(p) \gamma_5 + (1 - \frac{\xi}{2}) \bar{\Pi}_{cont}(p) - \frac{\xi}{2} \gamma_5  \bar{\Pi}_{cont}(p) \gamma_5,
\end{equation}
where $\Pi_{pole}(p) \simeq (1 - \frac{\xi}{2}) \bar{\Pi}_{pole}(p)$. Here last three terms are parts of the continuum (see detailed discussion in \cite{Kri2}) and the pole term takes the following form \cite{Coh1}

\begin{equation}
\Pi_{pole} (p) = -\lambda^{2}\frac{\slashed p + M^{*} + V\gamma_0}{2 E(p)(p^0 - E(p)} 
\end{equation}
Here $\lambda$ is the coupling of the nucleon interpolating field to the corresponding state of lowest energy, $M^*$ is the in-medium nucleon mass including the scalar part of the nucleon self-energy, $V$ is the vector part of the self-energy and $E(p)= \sqrt{p^2 + M{^{*}}^2} + V$.

According to the standard QCD SR procedure OPE and phenomenological representations of the correlator should be matched resulting in the following independent sum rules

\begin{equation}
-(1 - \frac{\xi}{2})\lambda^2 M^*\int d^4 p\frac{\omega(p)}{2E(p)[p^0 - E(p)]}\simeq (1 - \xi)\int d^4 \omega(p)[\bar{\Pi}^{s}_{OPE} - \bar{\Pi}^{s}_{cont}],
\end{equation}

\begin{equation}
-(1 + \frac{\xi}{2})\lambda^2\int d^4 p\frac{\omega(p)}{2E(p)[p^0 - E(p)]}\simeq \int d^4 \omega(p)[\bar{\Pi}^{p}_{OPE} - \bar{\Pi}^{p}_{cont}],
\end{equation}

\begin{equation}
-(1 + \frac{\xi}{2})\lambda^2 V\int d^4 p\frac{\omega(p)}{2E(p)[p^0 - E(p)]}\simeq \int d^4 \omega(p)[\bar{\Pi}^{u}_{OPE} - \bar{\Pi}^{u}_{cont}],
\end{equation}
where $\omega_(p)$ is some suitable weighting function to increase the overlap between the OPE and the phenomenological part of the QCD SR. 
Taking the ratio of these sum rules, one can get the required cancellation. The lessons to learn are that the way composite hadrons approach the chiral symmetry restoration can be different from one dictated by the order parameter of the lowest dimension and is basically defined by the  interplay between the different order parameters and their interactions with long-ranged Goldstone bosons.

Now we turn to the mirror assignment. In this case nucleons are genuine chiral partners related by the  commutation relations

\begin{equation}
[Q^{a}_5, \psi_1] = - \frac{\tau^a}{2}\gamma_5\psi_2, \qquad [Q^{a}_5, \psi_2] = - \frac{\tau^a}{2}\gamma_5\psi_1
\end{equation}

Under chiral rotations $\psi_1$ and $\psi_2$ are transformed into each other just like $\pi$ and $\sigma$ and thus belong to the same multiplet of $SU(2)_R \times SU(2)_L$. The pair of chiral partners is most often identified with physical nucleon and $N^*(1535)$ resonance although it is still an open issue. Sometimes the chiral partner of the nucleon is assumed to be the heavier resonance $N^*(1650)$. There also were suggestions to consider the hypothetical broad state with the mass centred around 1.2 GeV \cite{Pis} as the nucleon chiral partner but to our knowledge, the experimental search of such particle has not been performed so far. Using the expressions for the commutation relations in the mirror assignment one can obtain the chiral expansion for the nucleon correlators 

\begin{equation}
\Pi_N (p) =\bar{\Pi}_N (p) - \frac{\xi}{2} ( \bar{\Pi}_{N}(p) + \gamma_5\bar{\Pi}_{N^*}(p)\gamma_5),
\end{equation}
and the similar expression for the chiral partner

\begin{equation}
\Pi_{N^*} (p) =\bar{\Pi}_{N^*} (p) - \frac{\xi}{2} ( \bar{\Pi}_{N^*}(p) + \gamma_5\bar{\Pi}_{N}(p)\gamma_5).
\end{equation}
It is worth emphasizing that the mixing relations are very general and follow solely from the chiral symmetry constraints. If, for example, one utilises the QCD SR for the nucleon in mirror representation aiming at calculating the in-medium mass then one needs to include the mixing pattern written above. Any approach missing this point would at best be incomplete.

The correlator difference in the mirror assignment takes the form
\begin{equation}
\Pi_N (p)  - \Pi_{N^*} (p) = (\bar{\Pi}_N (p) - \bar{\Pi}_{N^*} (p))(1 - \frac{\xi}{2} ) + \frac{\xi}{2} (  \gamma_5(\bar{\Pi}_{N}(p) - \bar{\Pi}_{N^*} (p))\gamma_5),
\end{equation}

As one can see from the above equations in the mirror assignment nucleon and its chiral partner get mixed in the medium. Note that the first term in this expression also acts against CSR increasing the difference between correlators whereas the second one pushes the system towards it. Certainly, estimating the critical density using this mixing relations  does not seem realistic as near the point of CSR the $O(\xi^2)$ and higher order corrections as well contributions from heavy mesons, $\Delta$ - isobars and another nucleon resonances become important and they can't be obtained from chiral symmetry alone. Nevertheless, the mixing pattern of $N$ and $N^*$ suggest few possible scenarios of CSR in the parity doublet model:

1.  The masses of $N$ and $N^*$ become the same and equal to some chiral invariant mass $M_0$ as suggested in a number of effective models \cite{Pis,Har,Mish1}.

2.  One can also have a complete mixing of the correlators. Thus, they both exhibit peaks of equal strength, but chiral invariant masses of $N$ and $N^*$ are still different, implying that part of the $N/N^*$ masses is not related to the chiral symmetry, but rather to other mechanisms like for example density dependence of gluon condensate.

3.  Both correlators (or spectral functions) could be spreaded over the entire mass range. Thus the structural features of the correlators may be washed out and the discussion of the $N/N^*$ in-medium masses becomes meaningless.

One of the physical consequences of the parity mixing in the mirror assignment is that at finite density the Fermi sea for the chiral partner of the nucleon is formed effectively, after  pion corrections are taken into account.  This Fermi sea should be considered on the same ground as the nucleon one. It means that in the mirror assignment at sufficiently large densities nuclear matter becomes essentially asymmetric and contain  two mismatched  Fermi surfaces. Clearly in low density this effect is suppressed due to mass difference between nucleon and its chiral partner but the quantitative answer strongly depends on the assumed mass of the $N^*$ .  It also implies that in the mirror assignment the resulting nucleon density is smaller 
compared to that in the naive assignment. Another possible outcome of the mirror assignment in context of dense matter is that, when calculating for example the nucleon self-energy or response functions the states with the mass lower than one of $N^*$ like $\Delta (1235)$ and (probably) its chiral partner should also be taken into account. These mismatched Fermi surfaces open up a number of new options in the nuclear matter phase diagram including a possibility of the LOFF phase \cite{Loff} and Sarma phase \cite{Sar} which has been actively discussed in the other contexts.

One notes that cancellation of the $\rho m_\pi$ piece in the nucleon self-energy in matter also takes place in the mirror assignment as it follows from the first term of Eq.25 which is the same in both assignments in agreement with the chiral counting rules. The same holds for the nucleon chiral partner $N^*$.  However the in-medium nucleon effective mass, or more general, nucleon self-energy obtained for example in the framework of the QCD SR will be different in the naive and mirror representations. It would also be interesting and important to understand what part of nucleon self-energy is a chiral invariant and therefore does not depend on the assignment assumed. It is an open question at present.

It has already been mentioned that although the exited nucleon $N^*(1535)$ is the most popular candidate for the role of the nucleon chiral partner  it is not entirely clear issue at the moment. For example in the recent paper \cite{Har} the extended parity doublet model  has been suggested including four light nucleons $N(939)$, $N^*(1440)$, $N^*(1535)$, and $N^*(1650)$.
In this case the mixing pattern takes place among four chiral partners, related by some mixing coefficients. Clearly the main question here is whether such kind of models can incorporate a well established nuclear phenomenology. At the moment there are indications that the parity doublet model can successfully describe a number of nuclear observable although the problem is far from being settled. The other open issue concerns the value of $M_0$. Some authors \cite{Pis} advocate the value of chiral invariant nucleon mass $M_0$ being as low as 300-400 MeV whereas the others \cite{Har1} argue in favour of much higher value of $M_0$.

\begin{center}
2. IN-MEDIUM QUARK CONDENSATES.
\end{center}
As a byproduct of the developed approach we estimate the pion corrections to the in-medium two-quark condensate.
The vacuum quark condensate $<0|\bar{q}q|0> = <\bar{q}q>_0$ is an order parameter of QCD and manifests spontaneously broken chiral symmetry of QCD. Its in-medium counterpart $<\bar{q}q>_\rho$ can be defined as

\begin{equation}
<\bar{q}q(x)>_\rho = i Tr S(0,x)_\rho,
\end{equation}
where $S(0,x)_\rho$ stands for the quark propagator at finite density, which can be represented as a correlator of two quark fields

\begin{equation}
S(0,x)_\rho = <\psi|T\{q(x) \bar {q}(0)\}|\psi>.
\end{equation}
The spinor and color structure of the general matrix element can be projected out as follows

\begin{equation}
<|q_{a}(x) \bar{q}_{b}(0)|>_\rho = - \frac{\delta_{ab}}{12}\left(<|q(x) \bar {q}(0)|>_\rho + <|q(x)\gamma_\mu \bar {q}(0)|>_{\rho} \gamma^{\mu}\right)
\end{equation}
At vanishingly small distances and zero density the first term corresponds to standard scalar condensate which can be be extracted from the GOR relation (for example). The second term  ( "vector condensate") is obviously nonzero only at finite baryon density.

 With an increasing  density/temperature the value of the scalar condensate is supposed to decrease, thus signalling the tendency toward chiral symmetry restoration. Whereas the lattice calculations convincingly show that at finite temperature matter exhibits a  crossover transition to the chirally symmetric phase the same type of analysis at finite density is still hampered by a notorious sign problem. Therefore in the latter case one needs to rely on semianalytic approaches. In the leading order in density the drop of the quark condensate can quantitatively be described as follows

\begin{equation}
<\bar{q}q>_N = <\bar{q}q>_0 (1 - \frac{\rho \sigma_{\pi N}}{f^2_\pi m^2_\pi}).
\end{equation}

The leading finite density correction is model independent \cite{Dru}.  The next term is  proportional to $\frac{\partial}{\partial m^2_\pi}E_{int}(m_\pi, \rho)$ and describes the higher order corrections resulting  from the nucleon-nucleon interactions. Clearly the quantitative estimation of the higher order terms is model dependent. We note, however that all calculations indicate that the higher order corrections to the linear density approximation are rather small up to nuclear saturation densities and they grow when the nuclear density is increased. Still the quantitative estimates depend on the model assumed for the nuclear interactions. Our goal is to calculate the higher order corrections coming from the long range part of pion cloud using the above outlined technique, which is not based on any particular model adapted for the nucleon-nucleon interaction. Our estimates are therefore model independent but clearly less general compared to those where the effects of the heavy meson exchanges, three-body forces etc have been taken into account.

Using a standard commutation relation 
\begin{equation}
\left[Q^{a}_{5},q\right] = - \gamma_5\frac{\tau^a}{2}q,
\end{equation}

 one can get the corrections to the linear in density expression for the in-medium scalar quark condensate

\begin{equation}
<\bar{q}q>_\rho = <\bar{q}q>_N (1 - \xi), 
\end{equation}
 The $O(\xi^2)$ correction can be obtained by adding the next term with three pions in the expansion for the nuclear wave function $\psi_A$ and carrying on the same type of algebra but it leads to the negligible contributions in the region of densities where $\xi <<$ 1.

\begin{figure}
\begin{centering}
\includegraphics[width=13.5cm]{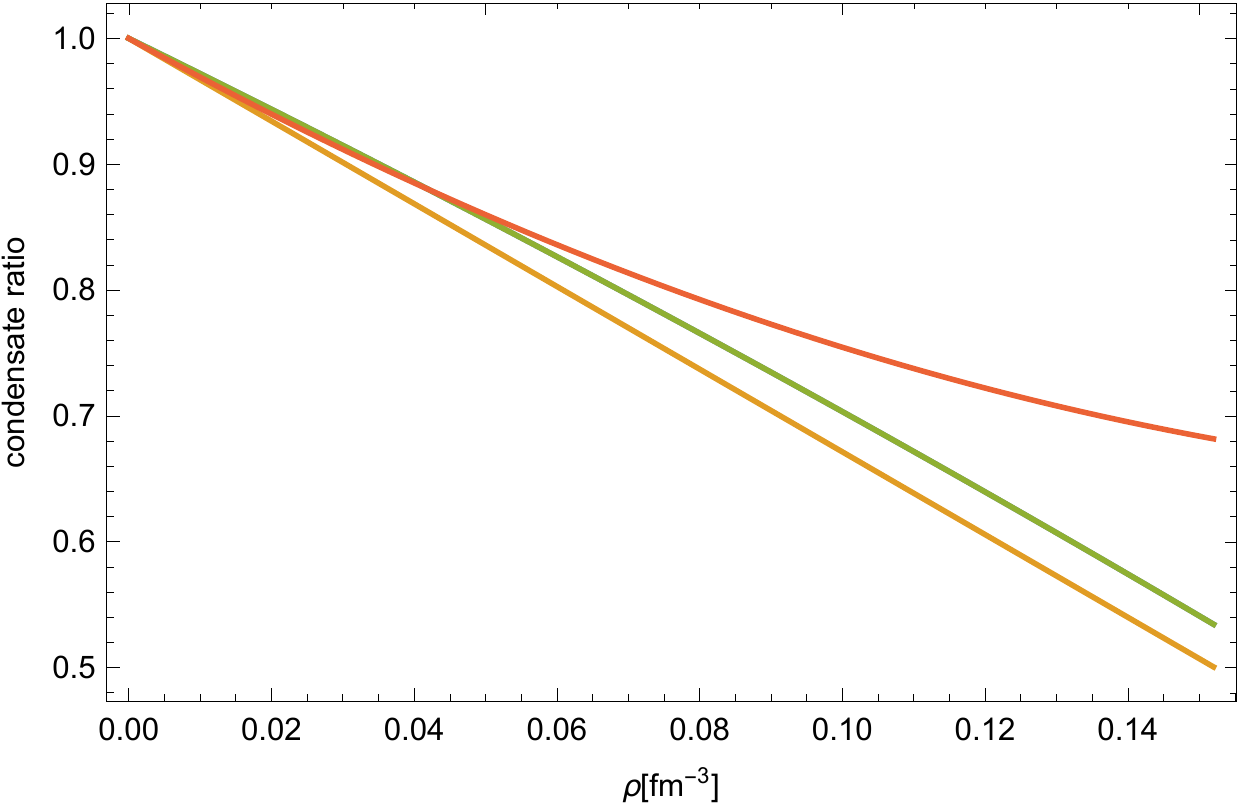}
\par\end{centering}
\vspace*{-5cm}
\vskip5cm
\caption{The ratio of the in-medium and vacuum quark condensates as a function of the nuclear density. The lower curve (brown online) is to the result  obtained in linear  density approximation, the middle curve (green online) corresponds to the case when soft pion corrections are taken into account and upper curve (red online) is the result from \cite{We} where all possible types of meson exchanges have been included.} . 
\end{figure}
One notes that the corrections to the lowest order approximation for the in-medium quark condensate have been studied in a number of papers \cite{We, Tsu} and led to quantitatively similar results indicating that higher order terms partially compensate the drop of the condensate found in the linear approximation resulting in the critical density of chiral symmetry restoration being much larger then the normal nuclear one. Some quantitative difference in the results is due to different models assumed for the nucleon-nucleon potential and numerical values of pion-nucleon sigma-term. We show on Fig.1 our (model independent) results and compare them with the results from Ref.\cite{We} where the in-medium quark condensate has been calculated using more realistic nuclear forces where the $1\pi$ and $2\pi$ exchanges as well as contribution from the $\Delta$ - isobar have been taken into account. One notes that the linear density approximation does a reasonably decent job in estimating the in-medium quark condensate almost up to nuclear matter saturation density which is not so trivial given the fact that it is far from correctly describing nuclear matter as a self-bound many-fermion system.

As one can see from Fig.1 there is a fairly good agreement between full calculations from \cite{We} and our estimate up to the density $\rho \simeq 0.07 - 0.08$ fm$^{-3}$. It implies that at relatively low densities the contributions from the long range pions constitute the significant part of the higher order corrections to the value of the in-medium quark condensate regardless of the model used for the NN interactions. One can conclude that the soft pion corrections act against CSR partially compensating the drop of the condensate following from the linear density approximation. At higher density the soft pion approximation becomes less trustworthy and cannot provide a stabilisation of the in medium condensate found in \cite{We, Tsu}. We note in passing that at densities bigger then $\rho = 0.2$ fm$^{-3}$ the calculations based on the potential models for the nucleon-nucleon interactions, either chiral or phenomenological, also become progressively less and less reliable.


As to "vector condensate" it is rather straightforward to show that the corresponding pion corrections for this condensate vanish. The same results holds for the related condensate $<q^+q>$. 

The next quantity we consider is the 4-quark condensate $<\bar{q}\Gamma q \bar{q}\Gamma q>$ with $\Gamma$ containing all possible spinor/flavour structures. Just like its 2-quark analogue the 4-quark condensate is the order parameter of the spontaneously broken chiral symmetry of QCD. However, unlike the $<\bar{q}q>$ condensate the four-quark one is rather poorly known, even in vacuum. The standard way of treating the 4-quark condensate used in the QCD sum rules based calculations is to utilize the factorisation approximation when the matrix element $<\bar{q}\Gamma q \bar{q}\Gamma q>$ is replaced with the product of two 2-quark condensates. Whereas this procedure can somehow be justified in the vacuum case using arguments based on the $\frac{1}{N}$ expansion, it is much less reliable or even wrong at finite density \cite{Kri2} as it turns out to be incompatible with the chiral symmetry requirements. As an example we will consider the  4-quark condensate  with the most trivial spinor/flavour structure $<\bar{q} q \bar{q} q>$. Following the above outlined procedure, now for the matrix element $<\psi|\bar{q} q \bar{q} q|\psi>$ one can get

\begin{equation}
<\bar{q} q \bar{q} q>_{\rho} \simeq
<\bar{q}q\bar{q}q>_N (1 - \xi) + \frac{2\xi}{3}<\bar{q}\gamma_{5} \frac{\vec{\tau}}{2}q \bar{q}\gamma_{5} \frac{\vec{\tau}}{2} q>_N,
\end{equation}
where the subscript N again means linear in nuclear density approximation. Just like in the 2-quark condensate case the pion corrections also result in moderate increase of the scalar part of the in-medium 4-quark condensate compared to the lowest order term although at different rate.  The novel feature is that the condensates with different types of spinor/flavour structures get mixed at finite density when soft pion corrections are taken into account. In the case considered here the scalar condensate gets mixed with the condensate containing two pseudoscalar diquarks. One notes that these two contributions partly compensate each other, although the quantitative estimate will be highly model dependent.  This result could to some extent be anticipated as the low lying scalar meson can be viewed as diquark-antidiquark state, i.e four-quark state. In dense medium the scalar meson can get mixed with its chiral partner, the pion which can also be interpreted as a superposition of many quark-antiquark states. Conversely, the expression for the $ <\bar{q}\gamma_{5} \tau_{\alpha}q \bar{q}\gamma_{5} \tau_{\alpha} q>$ condensate will also  include the contribution from the $<\bar{q}q\bar{q}q>$ structure. One notes that in the naive factorisation approximation (linear in density) the contribution from the second term vanishes. 
It is important to emphasize again that reliable estimate of the 4-quark condensate both in vacuum and at finite density/temperature is still open and difficult problem. It would be of a great interest to see the results of the lattice calculations for the 4-quark condensate to check the reliability of the factorization approximation. To our knowledge it has not been done so far.

\begin{center}  
3. MIXING PROPERTIES OF THE 3-POINT CORRELATION FUNCTIONS
\end{center}

Similar to the 2-point correlation functions 3-point correlators also get modified in the medium. Moreover, compared to the 2-point functions the 
mixing pattern of the 3-point correlators may have even more complicated structure. One notes that modification of the three-point functions in medium is to large extent unexplored issue so it is important to establish the general model independent structure of the correlators dictated by chiral symmetry. The general 3-point correlation function can be written as

\begin{equation}
\Pi_{J_1 J_2 J_3}(p)=-i \int d^4x \int d^4y\, exp (i p x) exp (i p y) <\psi|T\{J_{1}(x) J_{2}(y) J_{3}(0)|\psi>
\end{equation}
Using the soft pion approximation and  the general formula for the double commutator 

\begin{eqnarray}
\sum_{a,b}\left [Q^a_5,[Q^b_5, J_1 J_2 J_3]\right] = \sum_{a,b} ( J_1 J_2\left [Q^a_5,[Q^b_5,  J_3]\right] +  J_1\left [Q^a_5,[Q^b_5,  J_2] J_3\right] +\left [Q^a_5,[Q^b_5,  J_1]\right] J_2 J_3 \nonumber \\ + 2 [Q^a_5, J_1][Q^b_5,  J_2] J_3 +  2 [Q^a_5, J_1] J_2 [Q^b_5,  J_3] +  2 J_1 [Q^a_5, J_2][Q^b_5,  J_3])  
\end{eqnarray}
one can get the the expression for the pion corrections to the 3-point correlators calculated in the approximation of noninteracting nucleons. The concrete form of mixing depends on the commutation relations for the interpolating currents used in the correlator.

As an example, let's consider the process of the $\rho \rightarrow \pi\pi$ decay in nuclear matter which is described by the 3-point correlation function $\Pi_{\rho\pi\pi} (p)$.  This vertex plays a prominent role in hadron dynamics. The corresponding coupling constant in vacuum is fairly large $g_{\rho\pi\pi}\simeq$ 6 as well as couplings to nucleons and $\Delta$ isobars so one could expect a significant modification of this vertex in nuclear medium which in turn may be important in interpreting the dilepton data. These effects have been taken into account in Refs. \cite {Bron, Fri} where nuclear matter has been treated in the linear density approximation. The corresponding corrections were found to be significant. On top of that as will be shown below chiral symmetry prescribes a certain structure of the three-point correlator when the soft pion corrections are taken into account.

After some algebra, similar to that for the case of the 2-point correlators one can get the corresponding mixing relation

\begin{equation}
\Pi_{\rho\pi\pi} (p) =\bar{\Pi}_{\rho\pi\pi}  (p) -  (\alpha\bar{\Pi}_{\rho\pi\pi}(p) +  \beta\bar{\Pi}_{{\rho\sigma\sigma}}(p) + \gamma\bar{\Pi}_{{\sigma\pi A_1}}(p) ),
\end{equation}
Here $\bar{\Pi}$ denotes, as usual, the correlator calculated in the linear in density approximation and $\alpha, \beta, \gamma$ are some numerical factors whose exact values are not important in the given context. The expression for $\Pi_{\rho\pi\pi} (p)$ includes chiral partners of both pion and $\rho$-meson. One notes that even in the limit of noninteracting nucleons the three-point correlator is a very complicated object by itself. Moreover, the correlators $\bar{\Pi}_{{\rho\sigma\sigma}}(p)$ and $\bar{\Pi}_{{\sigma\pi A_1}}(p)$ are completely unexplored objects. Clearly, they are suppressed in vacuum but the situation is not so obvious for their in-medium analogues with the in-medium mass of the sigma meson being the main source of uncertainty.  Therefore, the above written mixing property should  be viewed as somewhat a formal statement merely illustrating the fact that chiral symmetry necessarily implies an existence of the contributions from chiral partners (if they are allowed by kinematics) which need to be taken into account in any realistic  theoretical model aiming at describing the $\rho \rightarrow \pi\pi$ decay in nuclear medium. 

Lets next consider the $\sigma\pi\pi$ vertex. The mixing relation takes the form 

\begin{equation}
\Pi_{\sigma\pi\pi} (p) =\bar{\Pi}_{\sigma\pi\pi}  (p) - (\delta\bar{\Pi}_{\sigma\sigma\sigma}(p) + \eta\bar{\Pi}_{{\sigma\pi\pi}}(p)),
\end{equation}

This relation may be of some importance in the context of an existence of a soft and narrow collective mode with the sigma meson quantum numbers when the chiral symmetry is (partially) restored \cite{Kun1}. As pointed out in 
\cite{Kun1} a possible experimental signal could be the enhancement of the spectral function and pion-pion cross section near the two-pion threshold. 

As the next example we will consider the correlator containing pion and two nucleon interpolating fields which can be used to extract the value of the in-medium $NN\pi$  coupling constant. This correlator (in vacuum) has been studied in the framework of the QCD sum rules in the Refs.\cite{Kri3, Hat1} where the calculated value of the $NN\pi$ coupling turned out to be very close to one extracted from the experimental data. We use the expression for the correlator where the time ordered product of two interpolating nucleon fields is sandwiched between the "vacuum" and the "one-pion" states. The general expression takes the form

\begin{equation}
\Pi_{NN\pi}(p)=-i \int d^4x \, exp (i p x) <\psi|T\{\eta(x) \bar{\eta}(0)\}|\pi\psi>
\end{equation}
Note that the external pion not need to be soft.
Again, after a straightforward algebra one can get the expression involving the commutator of the nucleon interpolating field and axial generator. Like in the two-point case we explore both naive and mirror assignments for the chiral properties of the nucleon. Let's consider the naive assignment first. Using the standard expressions for the nucleon commutators one can get

\begin{equation}
\Pi_{NN\pi}(p) = \bar{\Pi}_{NN\pi}  (p) - \frac{\xi }{2}(\bar{\Pi}_{NN\pi}(p) + \gamma_5\bar{\Pi}_{NN\pi}(p)\gamma_5),
\end{equation}

According to the standard QCD SR approach one should use Operator Product Expansion (OPE) on one hand and phenomenological description of the correlator on the other hand and then match them to extract physically relevant quantities. 

The OPE has a general structure

\begin{equation}
\int d^4x \, exp (i p x) T(\eta(x) \bar{\eta}(0)) = \sum_n C_n(q,\mu) O_n(\mu),
\end{equation}
where $C_n$ stands for the Wilson coefficient and $O_n$ denote the local operators, containing different combinations of quark and gluon fields and depending on renormalisation scale $\mu$ separating short and long range dynamics in $C_n$ and $O_n$ correspondingly.  Taking matrix element from vacuum to one pion state one can extract the pion-nucleon coupling constant in vacuum. If instead one takes the matrix elements between  the nucleon and nucleon plus one pion states one can determine the value of the pion-nucleon coupling constant in nuclear medium, applying for example the soft-pion theorem to the matrix elements $<N|O_n|N\pi>$ and utilising the values of the condensates, used to obtain the in-medium nucleon mass.  However, the practical implementation of this program is still hampered by the not very well known values of some condensates, especially the four-quark one.  On top of that, unlike the vacuum case the in-medium results may be sensitive to the choice of the interpolating field for the nucleon.
It is also worth mentioning that the QCD SR for the $NN\pi$ vertex formulated in this way are not completely independent but should rather be viewed as a chiral rotation of the two-point nucleon correlator \cite{Kri3}.

Now we turn to the mirror assignment. Using the corresponding commutation relations one can obtain the expression for the $\Pi_{N^{*}N^{*}\pi}(p)$ and $\Pi_{NN\pi}(p)$ correlation functions.

\begin{equation}
\Pi_{N^{*}N^{*}\pi}(p) = \bar{\Pi}_{N^{*}N^{*}\pi}  (p) - \frac{\xi }{2}(\bar{\Pi}_{N^{*}N{*}\pi}(p) + \gamma_5\bar{\Pi}_{NN\pi}(p)\gamma_5),
\end{equation}
and

\begin{equation}
\Pi_{NN\pi}(p) = \bar{\Pi}_{NN\pi}  (p) - \frac{\xi }{2}(\bar{\Pi}_{NN\pi}(p) + \gamma_5\bar{\Pi}_{N^{*}N{^*}\pi}(p)\gamma_5),
\end{equation}

As expected, the correlator includes the contributions from the chiral partner. It would be interesting to compare the values of the in-medium $\Pi_{\pi NN}$ coupling extracted from the correlators of the nucleons considered in both naive and mirror representations. However, it is by no means an easy task as the corresponding QCD SR analysis would require the information on a spectral functions of the chiral parter of nucleon. To our knowledge it has never been done so far.

The expressions for the off-diagonal correlator $\Pi_{NN^*\pi}$ in the mirror representation is

\begin{equation}
\Pi_{NN^{*}\pi}(p) = \bar{\Pi}_{NN^*\pi}(p) - \frac{\xi }{2}(\bar{\Pi}_{NN^{*}\pi}(p) + \gamma_5\bar{\Pi}_{NN{^*}\pi}(p)\gamma_5), .
\end{equation}
Similar expressions can be obtained for the correlators including for example electromagnetic current. It can be used for studying nucleon in-medium electromagnetic form-factors.

In order to understand which assignment is realized in nature one needs to come up with the experimental proposal which would distinguish between two  scenarios. There are few possible line of developments which can be pursued in this context. Some of then have been described in details in \cite{Oka1, Dirk1} so here we just summarize the main points. One possibility is to consider eta and pion production in the pion or photon induced reactions on the nucleon and use the interference effects to extract the relative sign of the $N N \pi$ and $N^*N^*\pi$ which is different in naive and mirror representations. Using nuclei instead may enhance the effect of interference but will also bring in further complications related to mechanisms not directly related to chiral symmetry like density dependent width of the $N^*$ resonance, the process of pion absorption or final state interaction \cite{Kri4}.

The other option is the photon induced production of the $\pi\eta$ pair. The idea is essentially the same as in the case of the single meson production and the main focus here is to look at angular distribution which is different in the mirror and naive assignments. Certainly the experimental set up for this reaction is much more difficult to realise in practice. 

\begin{center}  
5. SUMMARY 
\end{center}

We have discussed the mixing pattern between chiral partners for the meson and nucleon correlation functions at finite density which is prescribed by the chiral symmetry requirements. This mixing occurs regardless of the model used to calculate the correlation function at finite density. Such mixing suggests several possible scenarios of chiral symmetry restoration which does not necessarily imply that the masses of chiral partners become degenerate at the point of restoration. We mainly focused on the mixing pattern for the nucleon correlation functions where we used both naive and mirror assignment. Whereas in the naive assignment nucleon and its chiral partner transform independently in the mirror assignment the correlators  of nucleon and its chiral partner gets mixed thus leading to effective formation of the corresponding mismatched Fermi seas. It may potentially result in a number of interesting possibilities in the phase diagram.  We have also estimated the model independent contribution of the soft pions in the in-medium quark condensates. Whereas in the two-quark condensate case pion corrections lead to less pronounced decrease of the condensate compared to the lowest order estimates using noninteracting nucleon, in the four-quark condensate it leads to the mixing of the different types of the four-quark condensates.  

We have obtained the mixing patterns for the three-point correlation functions also exploring naive and mirror assignments for the nucleon interpolating field and pointed out few ways of checking which of those (naive or mirror) is realised in nature.

The directions of the future developments include revisiting QCD sum rules analysis of nuclear observables including symmetry energy, effective nucleon mass and binding energy as well as extension of the results to the combined finite temperature/density case.





\section{acknowledgement}
The author is grateful to M. Birse for a number of valuable discussions on this and related topics.


\begin{thebibliography}{16}

\bibitem{Lat}J. M. Lattimer and M. Prakash,  Phys.Rept., \textbf{621}, 127 (2016).

\bibitem{Sch} B. -J. Schaefer and J. Wambach, Nucl. Phys. \textbf{A757}, 479 (2005).

\bibitem{Mish}I. Mishustin, J. Bondorf and M. Rho, Nucl. Phys. \textbf{A555}, 215  (1993).

\bibitem{Moto}Y. Motohiro, Y. Kim and M. Harada.  Phys. Rev. \textbf{C 92}, 2, 025201 (2015).

\bibitem{Kri1}B. Krippa, Nucl. Phys. \textbf{A672}, 270  (2000)

\bibitem{Gas}J. Gasser, H. Leutwyler and M. E. Sainio,  Phys. Lett. \textbf{B253}, 152 (1991).

\bibitem{Lee}T. D. Lee and G. Wick, Phys. Rev. \textbf{D 9}, 2291 (1974).

\bibitem{Kun}C. E. Detar and T. Kunihiro, Phys. Rev. \textbf{D39}, 2805 (1989).

\bibitem{Hat}D. Jido, T. Hatsuda, and T. Kunihiro, Phys. Rev. \textbf{D57}, 4124 (1998).

\bibitem{Nem}Y.Nemoto et al, Phys. Rev. Lett\textbf{84}, 3252 (2000).

\bibitem{Sasa}C. Sasaki and I Mishustin,  Phys. Rev.  \textbf{C82}, 035204 (2010). 

\bibitem{Kri2} M. C. Birse and  B. Krippa,  Phys. Lett.\textbf{B381}, 397 (1996)

\bibitem{Wei}S. Weinberg,  Phys. Lett.\textbf{B251}, 288 (1990); Nucl. Phys. \textbf{B363}, 1 (1991).

\bibitem{Coh}S. H. Lee et al, Phys. Lett.\textbf{B348}, 263 (1995)

\bibitem{Ioff1}B. L. Ioffe, Nucl. Phys. \textbf{B188}, 679 (1981).

\bibitem{Coh1}T. D. Cohen, R. J. Furnstahl and D. K. Griegel, Phys. Rev.\textbf{C45}, 1881 (1992); T. D. Cohen, R. J. Furnstahl, D. K. Griegel, Xuemin Jin, Prog. Part. Nucl. Phys. \textbf{35} (1995) 221-298.

\bibitem{Pis}D. Zschiesche et al, Phys. Rev.\textbf{C75}, 055202 (2007).

\bibitem{Har}Y. Takeda, H. Abuki and M. Harada, Phys. Rev.\textbf{D97}, 094032 (2018).

\bibitem{Mish1}S. Benic, I. Mishustin and C. Sasaki, Phys. Rev.\textbf{D91}, 125034 (2015).

\bibitem{Loff}A. larkin and Y. Ovchinnikov, Zh. Eksp.Teor.Fiz.\textbf{47}, 1136 (1964); P. Filde and R. A. Ferell, Phys. Rev.\textbf{135}, A550 (1964).

\bibitem{Sar}G. Sarma, J. Phys. Chem.Solid \textbf{24}, 1029 (1963).

\bibitem{Har1}H. Nishihara and M. Harada,  Phys. Rev.\textbf{D92}, 054022 (2015).

\bibitem{Dru}E. G. Drukarev and E. M. Levin,  Nucl. Phys. \textbf{A511}, 679 (1990).

\bibitem{We}N. Kaiser and W. Weise,  Phys. Lett. \textbf{B671}, 25, (2009) 

\bibitem{Tsu}K. Tsushima et al., Euro Phys. J  \textbf{A31}, 626 (2007). 

\bibitem{Bron}W. Broniowski, W. Florkowski, and B. Hiller, Eur. Phys. J.\textbf{A7}, 287 (2000). 

\bibitem{Fri}B. Friman, Acta Phys. Polon., \textbf{B29}, 3195 (1998)

\bibitem{Kun1}D. Jido, T. Hatsuda and T. Kunihiro, Phys. Lett.\textbf{B670}, 109 (2008).

\bibitem{Kri3}B. Krippa and M. C. Birse, Phys. Lett.\textbf{B373}, 9 (1995); Phys. Rev.\textbf{C54}, 3240 (1996).

\bibitem{Hat1}S.Shiomi and T. Hatsuda, Nucl. Phys. \textbf{A594}, 294 (1995)

\bibitem{Oka1}D. Jido, M. Oka, and A. Hosaka, Prog. Theor. Phys. \textbf{106}, 873 (2001).

\bibitem{Dirk1}L. Olbrich et al, Phys. Rev. \textbf{D93}, 034021 (2016)
 
\bibitem{Kri4}  B. Krippa Phys. Rev.\textbf{C58}, 1333 (1998); B. Krippa and J. T. Londergan, Phys.Lett. \textbf{B286}, 216 (1992). 






\end{thebibliography}
\end{document}